\input{aipcheck}

\documentclass[
  ,draft            
  ]
  {aipproc}

\layoutstyle{8x11double}

\begin{document}

\newcommand{\lsi}   {LS~I~+61~303}
\newcommand{\mo}    {M$_{\odot}$}

\title{Gamma rays from compact binary systems}


\author{Josep M. Paredes}{
  address={Departament d'Astronomia i Meteorologia and Institut de Ci\`encies 
del Cosmos (ICC), Universitat de Barcelona (UB/IEEC), Mart\'{\i} i Franqu\`es 1,
08028 Barcelona, Spain\\E-mail: jmparedes@ub.edu}
}

\begin{abstract}
Some of the very high energy (VHE) gamma-ray sources detected with the 
modern generation of Cherenkov telescopes have been identified with 
previously known X-ray binary systems. These detections demonstrate the 
richness of non-thermal phenomena in compact galactic objects containing 
relativistic outflows or winds produced near black holes and neutron stars. 
Recently, the well-known microquasar Cygnus~X-3 seems to be associated with a gamma-ray source detected with {\it AGILE}.
Here I summarise the main observational results on gamma-ray emission 
from X-ray binaries, as well as some of the proposed scenarios to explain the 
production of VHE gamma-rays. 
\end{abstract}

\maketitle


\section{Introduction}

The detection of non-thermal X-ray emission from the jets of microquasars by the 
Chandra X-ray observatory \cite{Corbel02}, and
especially the detection of TeV gamma-rays by HESS and MAGIC from LS~5039 \cite{Aharonian05} and \lsi\ \cite{Albert06}
and the binary pulsar PSR~B1259$-$63 \cite{Aharonian05a}, 
provides a clear evidence of very efficient acceleration of particles
to multi-TeV energies in compact binary systems.  Under the conditions given in these objects, 
such high energies are hard to reach, mainly because of the expected strong radiation and magnetic fields present in these
systems. The study of these  binary systems at very high energies is of primary importance, since they are extremely
efficient accelerators that could shed new light and eventually force a revision of particle acceleration theory. The
astrophysical phenomena that could take place in these systems hardly compare with other TeV emitters. For instance, the
presence of strong photon fields could allow the study of photon-photon 
absorption and electromagnetic cascades on spatial and time 
scales that can be very small. In addition, the role played by the magnetic field is very important, since it directly 
affects the properties of the synchrotron radiation that likely ranges from radio to X-ray energies, and indirectly, 
the emission at higher energies, either suppressing electromagnetic cascades or modulating the TeV emission in
the case of an Inverse Compton (IC) origin. Hadronic processes, mainly proton-proton interaction, 
could also take place, 
generating gamma-ray
photons via neutral pion decay and electron-positron pairs via charged-pion decay. There are other effects to take into
account, like the role of the geometry in the interactions, 
namely photon-photon absorption and IC scattering due to the
orbital motion of the system. The particle injection mechanism could also vary periodically in the eccentric orbits of these
systems, adding new complexity but implying that information from the energy powering mechanism can also be extracted from
observations. Finally, it is worth to mention that these objects could also power extended X-ray and TeV emission.

\section{X-ray binaries}

An X-ray binary is a binary system containing a compact object, either a
neutron star or a stellar mass black hole, that emits X-rays as a result of a 
process of accretion of matter from the companion star. Several scenarios 
have been proposed to explain this X-ray emission, depending on the nature of 
the compact object, its magnetic field in the case of a neutron star, and the 
geometry of the accretion flow. The accreted matter is accelerated to
relativistic speeds, transforming potential energy provided by the intense 
gravitational field of the compact object into kinetic energy. Assuming that 
this kinetic energy is finally radiated, the accretion luminosity can be
computed, finding that this mechanism provides a very efficient source of
energy, which much higher efficiency than that for nuclear reactions.

In High Mass X-ray Binaries (HMXBs) the donor star is an O or B early type
star of mass in the range $\sim8$--$20$~M$_{\odot}$ and typical orbital
periods of several days. Most of HMXBs belong to two subgroups:
systems containing a B star with emission lines (Be stars), and systems
containing a supergiant (SG) O or B star.
In the first case, the Be stars do
not fill their Roche lobe, and accretion onto the compact object is produced
via mass transfer through a decretion disc. Most of these systems are
transient X-ray sources during periastron passage. In the second case, OB SG
stars, the mass transfer is due to a strong stellar wind and/or to Roche lobe
overflow. The X-ray emission is persistent, and large variability is common.
The most recent catalogue of HMXBs was compiled by \cite{Liu06},
and contains 114 sources in the Galaxy.

In Low Mass X-ray Binaries (LMXBs) the donor has a spectral type later than B,
and a mass $\leq2$~M$_{\odot}$. The orbital periods are in
the range 0.2--400 hours, with typical values $<24$ hours. The orbits are
usually circular, and mass transfer is due to Roche lobe overflow. Most
LMXBs are transients, probably as a result of an instability in the accretion
disc or a mass ejection episode from the companion. The typical ratio between
X-ray to optical luminosity is in the range $L_{\rm X}/L_{\rm
opt}\simeq 100$--$1000$, and the optical emission is dominated by X-ray heating of
the accretion disc and the companion star. The most recent catalogue of LMXBs was 
compiled by \cite{Liu07}, and contains 186 sources in the Galaxy.

Several X-ray binaries have been detected at radio wavelengths
with flux densities $\geq0.1$--$1$~mJy. The flux densities detected are
produced in small angular scales, which rules out a thermal emission
mechanism. The most efficient known mechanism for production of intense radio
emission from astronomical sources is the synchrotron emission mechanism, in
which highly relativistic electrons interacting with magnetic fields produce
intense radio emission that tends to be linearly polarized. The observed radio
emission can be explained by assuming a population of non-thermal
relativistic electrons, usually with a power-law energy distribution,
interacting with magnetic fields.

There are 9 radio emitting HMXBs and 55 radio emitting LMXBs. The 9 radio emitting HMXBs include 6
persistent and 3 transient sources, while among the 55 radio emitting LMXBs
we find 18 persistent and 37 transient sources. The difference between the
persistent and transient behavior clearly depends on the mass of the donor.
Although the division of X-ray binaries in HMXBs and
LMXBs is useful for the study of binary evolution, it is probably not
important for the study of the radio emission in these systems, where the main 
aspect seems to be the presence of an inner accretion disc capable
of producing radio jets. 
  
\section{Cygnus X-3: A new gamma-ray binary system?}

Cygnus X-3 is among the most intensively studied microquasars in the Galaxy. 
The system is a high-mass X-ray binary with a WN Wolf-Rayet companion star 
(see e.g. \cite{Fender99}, and references therein) seen through a
high interstellar absorption (A$_{V}\geq$ 10 mag) that renders the optical 
counterpart undetectable in the visual domain.

\begin{figure}
\resizebox{1.0\textwidth}{!} 
{\includegraphics{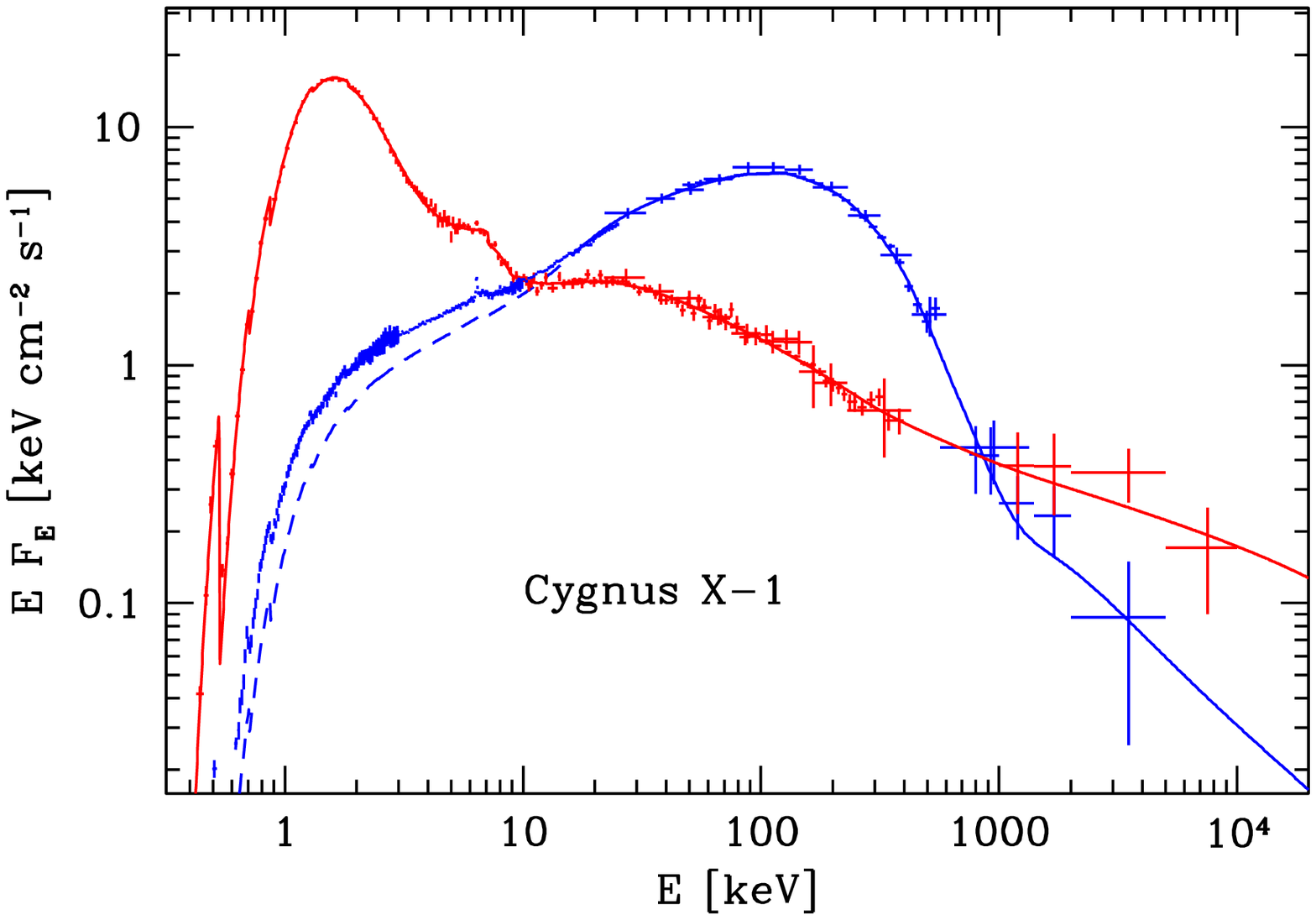}\hspace{1cm}\includegraphics{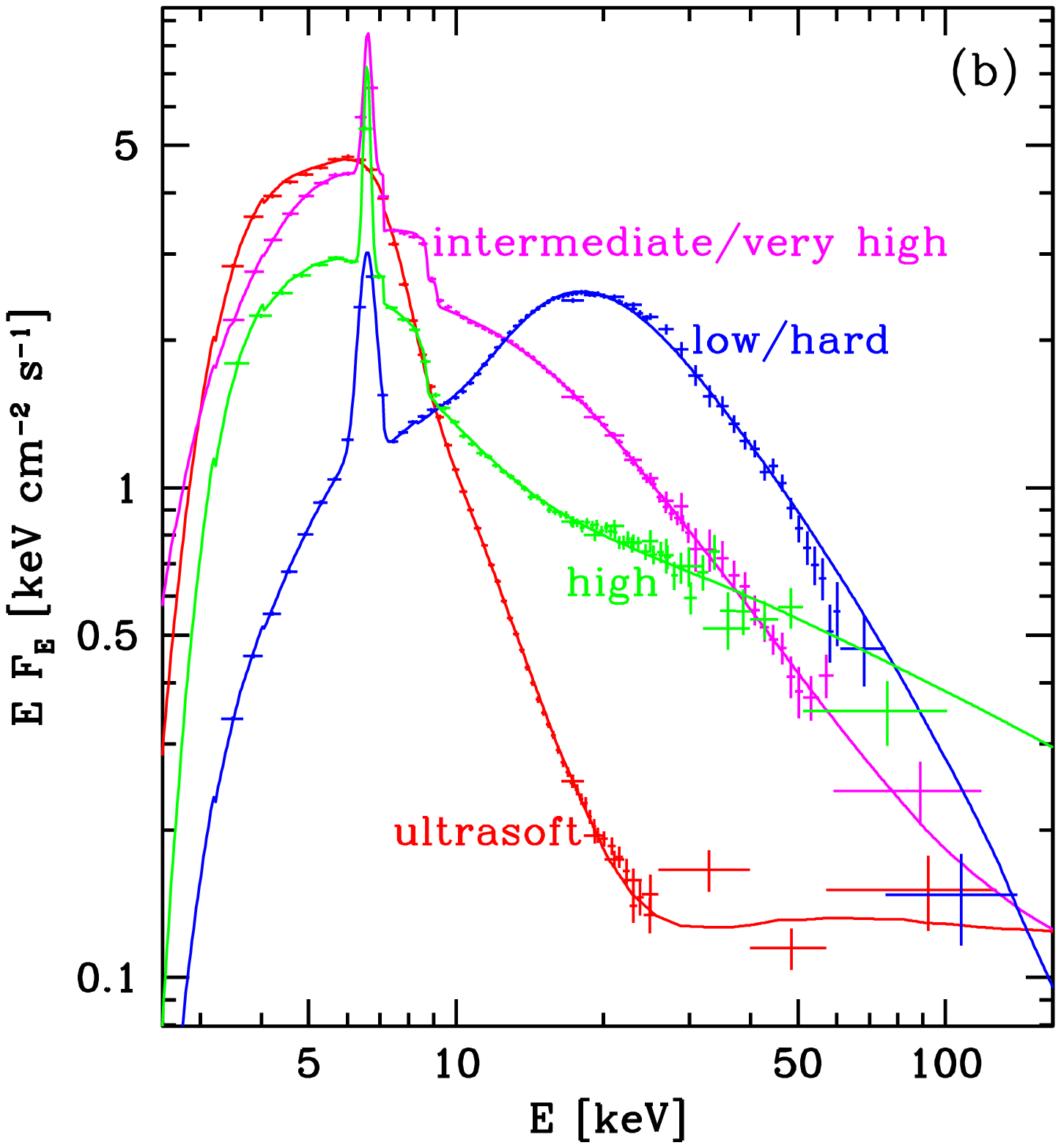}}
  \caption{Spectra of Cygnus~X-1 in the hard and soft state (left) 
and Cygnus~X-3 (right). Figure from Cygnus~X-1 is taken from~\cite{McConnell02} and Figure from Cygnus~X-3 is taken from~\cite{Zdziarski04}}
\label{cx1-cx3}
\end{figure}

The X-ray spectral states of Cyg X-3 closely correspond to the canonical 
X-ray states of BHBs although there are exceptions. One of these exceptions is
that the high-energy break in the hard state occurs in Cyg X-3 at $\sim$20 keV
whereas it is at $\geq$ 100 keV in other objects. Another exception is the 
strong absorption. We can see an example of this in Fig.~\ref{cx1-cx3}, where 
we can compare the spectral behaviour of Cygnus~X-1 in its hard and soft states
with that of Cygnus~X-3.

\begin{figure}
\resizebox{1.0\textwidth}{!}
   {\includegraphics[angle=270]{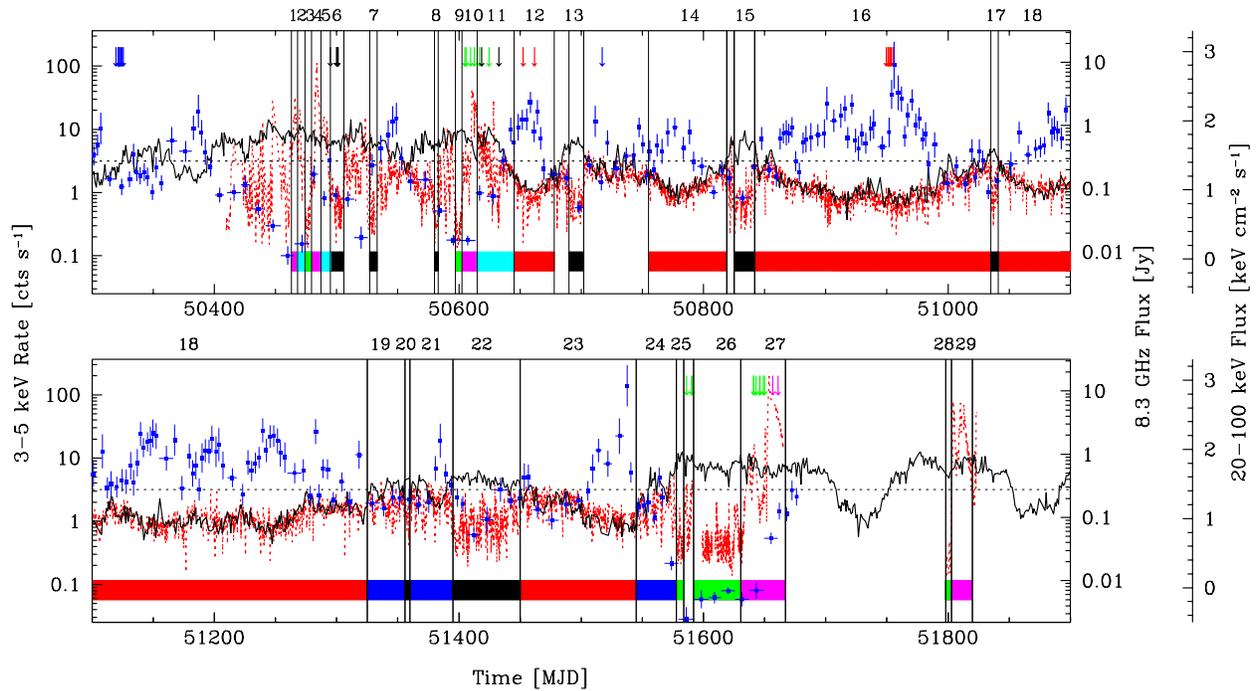}}
 \caption{Multi-wavelength monitoring of Cygnus X-3 spanning 4,4 yr, including
 lightcurves obtained with the {\it RXTE} ASM 3 -- 5 keV (black solid curve), BATSE 20 -- 100 keV 
(blue squares), and GBI 8.3 GHz (red dotted curve). 
The horizontal dotted line represents the transition level at which the radio/soft X-ray 
correlation changes its character. The vertical lines divide
the light curves into intervals of different activity types. Figure taken 
from~\cite{Szostek08}}
\label{multi}
\end{figure}

The system often undergoes strong radio outbursts, one or two times per year, 
with flux density increments of almost three orders or magnitude above the 
normal quiescent level, of $\sim$0.1 Jy at cm wavelengths. The first of such 
event observed was the historic radio outburst extensively described by 
\cite{Gregory72} and subsequent papers. Collimated
relativistic jets from this microquasar were reported soon after some of these 
flaring episodes, flowing away in the
North-South direction (see e.g. \cite{Marti01}; \cite{Miller-Jones04}).
Cygnus~X-3 appears superposed onto a diffuse radio emission with apparent nonthermal index with an angular size of a few arc-minutes
extending South and South-West from it \cite{Sanchez08}. It has been suggested the possibility that such an extended emission could be physically
associated to Cygnus~X-3.
Strong radio flares occur only when the source is in the soft state. An example of this is shown in Fig.~\ref{multi}, where 
{\it RXTE} ASM 3 -- 5 keV, BATSE 20 -- 100 keV and GBI 8.3 GHz observations of Cygnus~X-3 along more than four years are 
shown (\cite{Szostek08}).

According to \cite{Szostek08}, if the non-thermal electrons responsible for  either the hard X-ray tails  or the radio
emission during major flares were accelerated  to high enough energies then detectable emission in the gamma-ray range would
be possible. In consequence, given that major radio flares indicates the presence of hard X-ray tails, GeV and TeV emission
should be searched for during those radio flares, although the strong present photon fields could absorbe this radiation if originated inside the system.


Recently, a giant radio flare from the microquasar Cygnus X-3 was detected 
with the RATAN-600 radio telescope on 18 April 2008 \cite{Trushkin08}. This 
powerful flare was observed simultaneously at seven frequencies, covering the 
range from 1 to 30 GHz. The maximum flux density measured was 16.2 Jy at 11.2 
GHz and the spectrum was optically thick at frequencies lower than 4.8 GHz.
This radio outburst activity of Cygnus~X-3 corresponded to an X-ray soft state 
according the data from {\it Swift}/BAT (15--55 keV) and {\it RXTE}. 

The Italian {\it AGILE} gamma-ray satellite collected data from the Cygnus region at several epochs during 
the period 2007--2008, showing a > 100~MeV integrated emission possibly associated with Cygnus~X-3, 
and with an slight increase of the emission on 18 April 2008 \cite{Giuliani08}. Although the goodness 
of the association of the emission to Cygnus~X-3 must be confirmed, these results point to the possibility 
of Cygnus~X-3 being a new high energy gamma-ray source. At TeV energies, Cygnus~X-3 has not yet been detected with the 
new generation of Cherenkov telescopes. In the past there were some claims of its detection (\cite{Chadwick85}) 
but have not been considered valid because of instrumental limitations at these epochs.

\section{The VHE binaries}

At present,
there are four X-ray binaries that have been detected at TeV energies.
Three of them, PSR~B1259$-$63, LS~I~+61~303 and LS~5039 have been detected in several 
parts of their orbits and show a variable TeV emission. The other source, Cygnus~X-1, has been detected once during a 
flare. LS~I~+61~303 shares with LS~5039 the quality of being the only two known
high-energy emitting X-ray binaries that are spatially coincident with sources
above 100~MeV listed in the Third EGRET catalog \cite{Hartman99}. Some properties of these systems 
are summarized in Table~\ref{table}, and these sources are individually described below.

\begin{table}[]
\begin{tabular}{lccccc}
\hline
{Parameters} & {PSR~B1259$-$63} & {LS~I~+61~303} & {LS~5039} & {Cygnus~X-1} & {Cygnus~X-3} \\
\hline
System Type  & B2Ve+NS & B0Ve+NS?  & O6.5V+BH? & O9.7Iab+ BH & WN$_{\rm e}$+ BH?\\[3pt]
Distance (kpc) & 1.5 & 2.0$\pm$0.2             & 2.5$\pm$0.5 & 2.2$\pm$0.2 & $\sim$9 \\[3pt]
Orbital Period (d)  & 1237 & 26.5  & 3.90603$\pm$0.00017 & 5.6 & 0.2 \\[3pt]
$M_{\rm compact}$ (M$_{\odot}$)  & 1.4   & 1--4  & 1.4--5  & 20$\pm$5 & --\\[3pt]   
Eccentricity & 0.87 & 0.72      & 0.35$\pm$0.04 & $\sim$ 0 & $\sim$ 0 \\[3pt]
Inclination & 36 &  $30\pm 20$     & 20? & $33\pm 5$ & -- \\[3pt]
Periastron-apastron (AU) & 0.7--10 &  0.1--0.7  & 0.1--0.2 & 0.2 & --\\[3pt]

\hline
Physical properties & &   &  &  & \\[3pt]
\hline
Radio Structure (AU) & $\le$2000 & Jet-like (10--700)  & Jet-like (10--$10^{3}$)  & Jet
(40) + Ring & Jet $\sim 10^{4}$  \\[3pt]
$L_{\rm radio(0.1--100~GHz)}$ (erg s$^{-1}$) & (0.02--0.3)$\times10^{31}$ $^{\rm (*)}$ & 
(1--17)$\times10^{31} $ &$1\times10^{31}$   & $0.3\times10^{31}$ & $7\times10^{32}$ \\[3pt]
$L_{\rm X(1--10~keV)}$ (erg s$^{-1}$) & (0.3--6)$\times10^{33}$  & (3--9)$\times10^{33}$  & (5--50)$\times10^{33}$  & 
$1\times10^{37}$ & $(3.9-7.9)\times10^{37}$ \\[3pt]
$L_{\rm VHE}$ (erg s$^{-1}$) & $2.3\times10^{33}$ $^{\rm (a)}$  & $8\times10^{33}$ 
$^{\rm (a)}$  & $7.8\times10^{33}$ $^{\rm (b)}$ & $12\times10^{33}$ 
$^{\rm (a)}$  & -- \\[3pt]
$\Gamma_{\rm VHE}$ & $2.7\pm0.2$ & $2.6\pm0.2$  & $2.06\pm0.05$ & $3.2\pm0.6$ & -- \\[3pt]

\hline
Periodicity & &   &  &  & \\[3pt]
\hline
Radio &  48 ms and 3.4 yr  &  26.496 d and 4.6 yr & persistent & 5.6 d & persistent and strong outbursts \\[3pt]
Infrared &-- & 27.0$\pm$0.3 d  & variable & 5.6 d & -- \\[3pt]
Optical & --&  26.4$\pm$0.1 d& -- & 5.6 d & --\\[3pt]
X-ray & variable& 26.7$\pm$0.2 d  & variable & 5.6 d & 0.2 d \\[3pt]
$>$ 100 MeV & -- & variable  & variable ? & -- & variable ? \\[3pt]
$>$ 100 GeV & variable & 26.8$\pm$0.2 d  & 3.9078$\pm$0.0015 d& flare & -- \\[3pt]
\hline
{\small $^{\rm (*)}$ Unpulsed radio emission} \\
{\small $^{\rm (a)}$ 0.2 $<$E$<$ 10 TeV} \\
{\small $^{\rm (b)}$ Time averaged luminosity.} \\
\end{tabular}
\caption{The five X-ray binaries that are MeV and/or TeV emitters}
\label{table}
\end{table}

\subsection{PSR~B1259$-$63}

PSR B1259$-$63 / SS 2883 is the first variable galactic source of VHE gamma-rays. This is a binary system containing a B2Ve
donor and  a 47.7 ms radio pulsar orbiting it every 3.4 years in a very eccentric orbit with e = 0.87. During the orbital
phases where the neutron star is behind the circumstellar disk, its pulsed radio emission is not observed because of free-free
absorption. The radiation mechanisms and  interaction geometry in this pulsar/Be star system was studied in \cite{Tavani97}.
It was found that the observed high-energy emission from the PSR B1259$-$63 system is not compatible with accretion or
propeller-powered emission, whereas it is consistent with the shock-powered high-energy emission produced  by the
pulsar/outflow interaction. 

In \cite{Aharonian05a} it is reported the discovery of VHE gamma-ray emission of PSR B1259$-$63 system by HESS. The TeV emission
is detected when the neutron star is close to periastron or crosses the disk, and the flux varies significantly on timescales
of days.  The time-averaged energy spectrum (above 380\,GeV) can be fitted by a power law $F_0(E/1\,\rm TeV)^{-\Gamma}$ with
a photon index $\Gamma = 2.7\pm0.2_\mathrm{stat}\pm0.2_\mathrm{sys}$ and flux normalisation $F_0 = (1.3 \pm 0.1_\mathrm{stat}
\pm 0.3_\mathrm{sys}) \times 10^{-12}\,\rm TeV^{-1}\,\rm cm^{-2}\,\rm s^{-1}$ \cite{Aharonian05a}.

Different models have been recently proposed to try to explain these observations. In a hadronic scenario, the TeV 
light-curve, and radio/X-ray light-curves, can be produced by the collisions of high energy protons accelerated by the pulsar
wind and the circumstellar disk, being the VHE $\gamma$-rays produced in the decays of secondary $\pi^{0}$, while radio and
X-ray emission are synchrotron and IC emission produced by low-energy ($<$ 100 MeV) electrons from the decays of secondary
$\pi^{\pm}$ \cite{Neronov07}.  A very different model is presented in \cite{Khangulyan07}, where it is shown that the
TeV light curve can be explained by  IC scenarios of gamma-ray production. Moreover,  the Comptonization of the pulsar  wind
leads to the formation  of gamma-radiation with a line-type energy spectrum that should appear either at GeV or TeV energies
depending of the initial Lorentz factor of the wind  \cite{Khangulyan07}.

Very recent HESS observations \cite{Kerschhaggl08}, covering unexplored orbital phases prior to periastron, show a clear
pre-periastron detection. These results will allow a new insight to understand more clearly this system.

\subsection{LS~I~+61~303}

\lsi\ is a high mass X-ray binary that shows periodic non-thermal radio outbursts on average every $P_{\rm orb}$=26.4960~d
(\cite{Taylor82}, \cite{Gregory02}). The system is composed of a rapidly rotating early type B0\,Ve star with a stable
equatorial decretion disk and mass loss, and a compact object with a mass between 1 and 4 $M_{\odot}$ orbiting it every
$\sim$26.5~d in a highly eccentric orbit with $e=0.72$ \cite{Casares05a}. Spectral line radio observations give a distance of
2.0$\pm$0.2~kpc \cite{Frail91}. \cite{Massi04} reported the discovery of an extended jet-like and apparently precessing radio
emitting structure at angular extensions of 10--50~milliarcseconds. Due to the presence of (apparently relativistic) radio
emitting jets, \lsi\ was proposed to be a microquasar. However, recent VLBA images obtained during a full orbital cycle show
a rotating elongated morphology \cite{Dhawan06}, which may be consistent with a model based on the interaction between the
relativistic wind of a young non-accreting pulsar and the wind of the stellar companion \cite{Dubus06}.  

Possible evidence of an X-ray extended structure at a distance between 5" and 12" toward the north of LS~I~+61~303 has been found
at a significance level of 3.2 $\sigma$ \cite{Paredes07}.

\subsubsection{$Gamma$-ray source}

2CG~135+01 was one of the most prominent unidentified gamma-ray sources near the Galactic plane discovered by the {\it COS B}
satellite \cite{Hermsen77}. The {\it COS B}  error box contained \lsi\ and since its discovery as a variable radio source
\cite{Gregory78} was proposed to be associated with 2CG 135+01. EGRET observations of 2CG~135+01/3EG J0241+6103 showed that
the only likely source that is spatially coincident with the gamma-ray position is the radio source GT 0236+610/LS~I~+61~303
\cite{Kniffen97}. The instrument COMPTEL \cite{Schonfelder00} detected the source GRO~J0241+6119, being
the most likely counterpart LS~I~+61~303, although its emission in the range 1--30 MeV is possibly contaminated by the quasar
QSO~0241+622. Recently, {\it AGILE} has also detected LS~I~+61~303 \cite{Giuliani08}. At higher energies, the MAGIC \v{C}erenkov
telescope discovered \lsi\ at very high energy gamma rays ($E_\gamma>$100~GeV; \cite{Albert06}).  Recent observations by the
MAGIC and VERITAS collaborations clearly indicate the existence of an orbital TeV variability in \lsi\ (\cite{Sidro07} and
\cite{Acciari08}, respectively). The MAGIC measurements showed that the maximum flux corresponded to about 16\% of that of
the Crab Nebula, and was detected around phase 0.6 with 8.7$\sigma$ of significance. The spectrum derived from MAGIC data
between 200~GeV and 4~TeV at orbital phases between 0.4 and 0.7 is fitted by a power law function: $F_\gamma = (2.7 \pm 0.4
\pm 0.8) \times 10^{-12} (E/{\rm TeV})^{-2.6 \pm 0.2 \pm 0.2}$ cm$^{-2}$~s$^{-1}$~TeV$^{-1}$, with the errors quoted being
statistical and systematic, respectively \cite{Albert06}.

\subsubsection{Multiwavelength periodicity}

\begin{figure}
\resizebox{0.45\textwidth}{!}
{\includegraphics{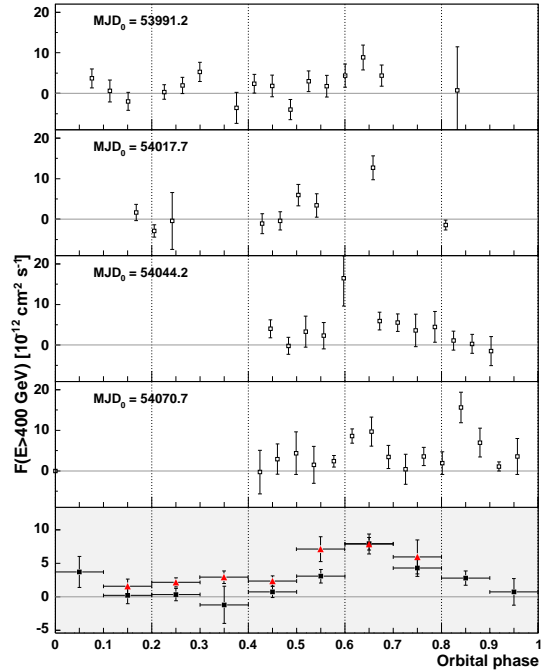}}
  \caption{VHE gamma-ray flux of LS~I +61~303 as
a function of the orbital phase. The four upper panels correspond to the four observed orbital cycles and the lowermost panel correspond to the averaged for the entire observation time. The previous published \cite{Albert06} averaged fluxes per phasebin are shown in red. Vertical error bars include 1$\sigma$ statistical error. Figure taken 
from~\cite{Albert08}.}
\label{lsi-MAGIC}
\end{figure}

One of the most unusual aspects of its radio emission is the fact that it exhibits two periodicities: periodic nonthermal
outbursts every 26.496 day \cite{Taylor82} and a 1667 day ($\sim$4.6 years) modulation of the outburst peak flux
(\cite{Paredes87}, \cite{Gregory02}). The 26.5 day periodicity corresponds to the orbital period of the binary system
\cite{Hutchings81}. This periodicity has also been detected in the optical  \cite{Mendelson89} \cite{Mendelson94} and the
infrared domains \cite{Paredes94}, in soft X-rays \cite{Paredes97} and in the H$\alpha$ emission line \cite{Zamanov99}. The
$\sim$4.6 year modulation has been observed as well in the H$\alpha$ emission line \cite{Zamanov99}. Recently, periodic very
high energy gamma-ray emission from LS~I~+61~303 has been observed with the MAGIC telescope (\cite{Albert08},
\cite{Sidro08}). The gamma-ray flux ($E >$ 400~GeV) is plotted in  Fig.~\ref{lsi-MAGIC} as a function of the orbital phase. 

EGRET observations of 3EG J0241+6103 shows variability on short (days) 
and long (months) timescales \cite{Tavani98}. {\it AGILE} has 
detected recently this source at $>$ 100 MeV \cite{Giuliani08}. 
Future deep observations spanning several months could give hints
of periodicity at these energies.  

\subsection{LS~5039}

LS~5039/RX~J1826.2$-$1450 was first identified as a new massive X-ray binary by \cite{Motch97}. The first radio
detection was reported by \cite{Marti98} using the Very Large Array. The radio emission is persistent, non-thermal
and variable but no strong radio outbursts or periodic variability have been detected so far (\cite{Ribo99}, \cite{Ribo02}). The detection of an elongated radio structure, interpreted as relativistic jets, was only possible when the source was
observed at millisecond-arc scales with the Very Long Baseline Array \cite{Paredes00}. 

In the optical band LS~5039 appears as a bright $V$=11.2, O6.5V((f)) star showing little variability on timescales of months
to years \cite{Clark01}. Variations of $\sim$0.4~mag have been reported in the infrared ($H$ and $K$ bands) but no
obvious mechanisms for such variability have been proposed \cite{Clark01}. The orbit of LS~5039 was first studied by
\cite{McSwain01}, and more recently by \cite{Casares05b} who found an orbital period of $P_{\rm orb}=3.9$ day and an
eccentricity of $e=0.35$. The mass of the compact object is still unknown, ranging between 1.4 and 5 $~M_{\odot}$, depending
on the binary system inclination.

\subsubsection{$Gamma$-ray source}

\begin{figure}
\resizebox{0.45\textwidth}{!}
   {\includegraphics{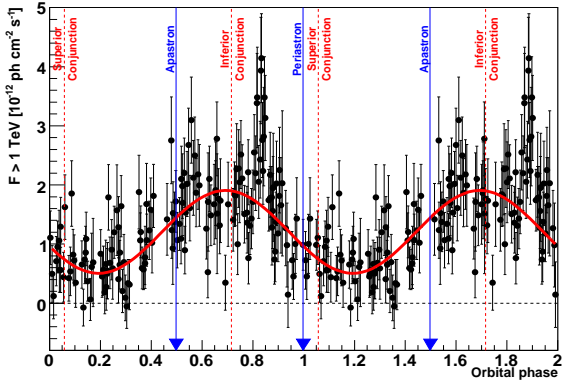}}
  \caption{HESS integral $\gamma$-ray flux of LS~5039 as a function of orbital 
phase \cite{Aharonian06}.}
\label{ls}
\end{figure}

The discovery of the bipolar radio structure, and the fact that LS~5039 was the only source in the field of the EGRET
source 3EG~J1824$-$1514 showing X-ray and radio emission, allowed to propose the physical association of both sources
\cite{Paredes00}.  LS~5039 is also one of the possible counterparts of  GRO~J1823$-$12, which is among the strongest
COMPTEL (1--30 MeV) sources. The source region, detected at high significance level, contains several possible counterparts,
being LS~5039 one of them \cite{Collmar04}. LS~5039 has been detected at very high-energy gamma-rays \cite{Aharonian05}, which gives
strong support to the proposed association with the EGRET and COMPTEL source. The TeV emission shows a periodicity identical
to the orbital period \cite{Aharonian06} (see Fig.~\ref{ls}). It can be seen that the flux is maximum at inferior conjunction
of the compact object. This   suggests that photon-photon absorption (e$^{+}$$-$e$^{-}$ pair production on stellar UV
photons), which has an angle dependent cross-section plays a major role but the flux should be zero at periastron and
superior conjunction, and this is not the case, while the spectrum shows strong variability, but not at 200 GeV as predicted by absorption
models (\cite{Dubus06}, \cite{Boettcher 2007}).

The differential photon energy spectrum is variable with orbital phase. During the phases of the compact object inferior
conjunction the spectrum is consistent with a hard power-law where $\Gamma_{\rm VHE}$ = $1.85\pm0.06_{\rm stat}\pm0.1_{\rm
syst}$ with exponential cutoff at $E_0$ = $8.7\pm2.0$ TeV. At the superior conjunction phases, the spectrum is consistent
with a relatively steep ($\Gamma_{\rm VHE}$ = $2.53\pm0.07_{\rm stat}\pm0.1_{\rm syst}$) pure power-law (0.2 to 10 TeV). The
HE/VHE emission is basically interpreted as the result of inverse Compton upscattering of stellar UV photons by relativistic
electrons.

\subsubsection{Changing radio morphology}

\begin{figure}
\resizebox{1.0\textwidth}{!}
   {\includegraphics{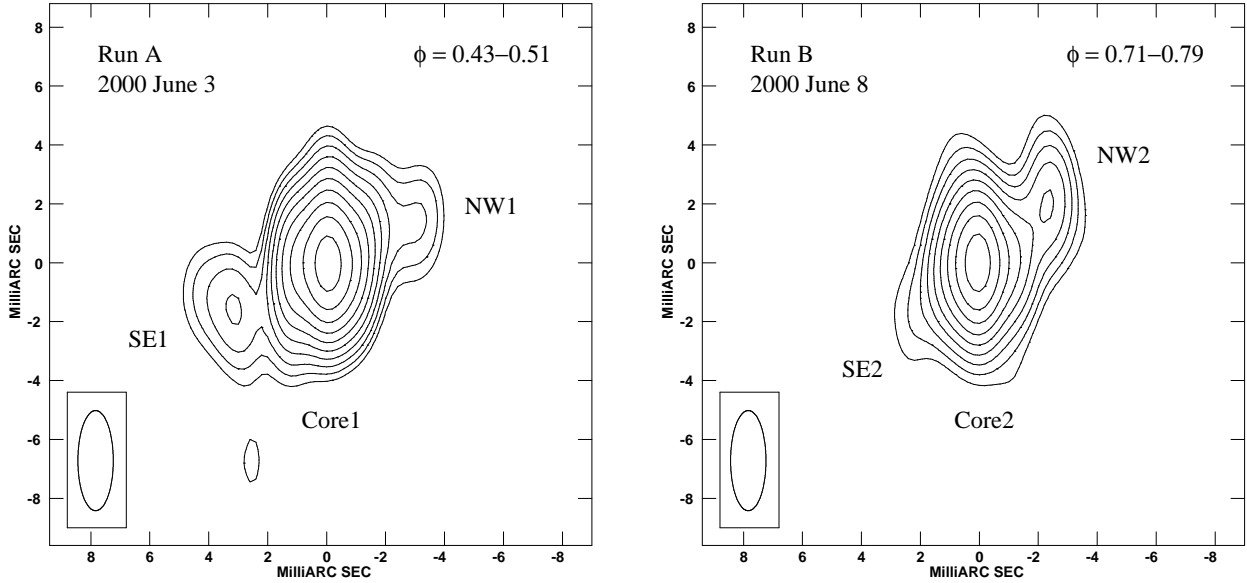}}
  \caption{VLBA maps of LS~5039 at different orbital phases (\cite{Ribo08}).}
\label{lsvlbi}
\end{figure}

High resolution radio observations of LS~5039 with Very Long Baseline 
Interferometry (VLBI) can bring valuable information to advance in our 
knowledge of this source. The detection of morphological and astrometric
changes can be useful to disantangle between the two possible scenarios. 
Recent VLBA+VLA images of LS~5039 \cite{Ribo08}\cite{Moldon07}obtained
at two different orbital phases show a changing morphology 
(see Fig.~\ref{lsvlbi}).
There is extended
radio emission that appears nearly symmetric for run A and clearly 
asymmetric for run B, with a small change of $\sim$10$^\circ$ in its position angle. New VLBI observations are required to obtain morphological information
along the orbit.

In any case, precise phase-referenced VLBI observations covering
a whole orbital cycle are necessary to trace possible periodic displacements
of the peak position and to obtain morphological information
along the orbit. These might put some constraints on the nature
of the powering source in this gamma-ray binary.  

\subsubsection{Runaway system}

\begin{figure}
\resizebox{0.5\textwidth}{!}
   {\includegraphics{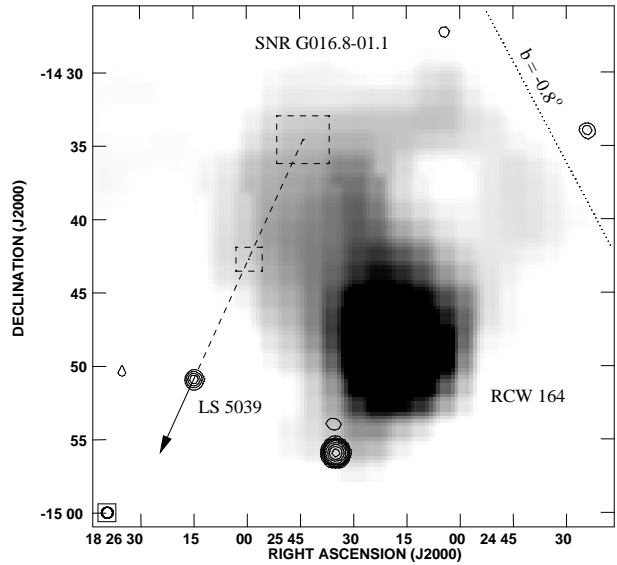}}
  \caption{Radio map of LS~5039 showing the HII-region RCW~164 (the stronger source in the field) and the shell-like SNR G016.8$-$01.1. The arrow marks the proper motion sense as was found in \cite{Ribo02t}. A new proper motion has been obtained recently \cite{Moldon07}.}
\label{lsmovpro}
\end{figure}

In \cite{Ribo02}, the origin of the gamma-ray binary LS~5039 was explored by  means of VLBI observations, to unveil if it is
coming from the SNR G016.8$-$01.1. It was shown that LS~5039 is a runaway system moving away from the Galactic plane with a
total systemic velocity of $\sim$ 150~km~s$^{-1}$ and a component perpendicular to the Galactic plane larger than
100~km~s$^{-1}$ (\cite{Ribo02}, \cite{McSwain02}). The escaping velocity of this system from its local environment may be
the result of the supernova explosion which created the compact object in this binary system. According to the computed
trajectory, LS~5039 could reach a galactic latitude of $-$12$^{\circ}$ 
before the donor star evolves and the X-ray source disappears (see Fig.~\ref{lsmovpro}). However, from the kinematical point of view, it was
not possible to clearly confirm nor reject the association between LS~5039 and SNR G016.8$-$01.1. 

Recently, a new proper motion that
is compatible with the center of  SNR G016.8$-$01.1, has been calculated using high accuracy interferometric observations from 1998 to 2007 \cite{Moldon07}.

\subsection{Cygnus~X-1}

Cygnus~X-1 is the first binary system where dynamic evidence for a BH was found \cite{Gies86}. According to the most recent
estimates, the BH mass is 20$\pm$5~$M_{\odot}$, while the O9.7\,Iab supergiant companion has a mass of 40$\pm$10~$M_{\odot}$
\cite{Ziolkowski05}. The orbit of the system is circular, with a period of 5.6 days and an inclination of 33$\pm$5$^{\circ}$
\cite{Gies86}. Located at 2.15$\pm$0.20~kpc (3$\sigma$ error; see \cite{Ziolkowski05} and references therein),  Cygnus~X-1 is
the brightest persistent HMXB in the Galaxy, radiating a maximum X-ray luminosity of a few times 10$^{37}$ erg~s$^{-1}$ in
the 1--10~keV range. 

The source displays the typical low/hard and high/soft states of accreting BH
binaries, spending most of the time, currently about 65\%, in the low/hard
X-ray state \cite{Wilms06}. Steady compact jets are produced in BH
binaries in this state, when the inner radius of the disk is thought to be
truncated, while in the high/soft state the jet is quenched \cite{Fender04}. 
This is also the case for Cygnus~X-1, which displays a $\sim$15~mJy and
flat spectrum relativistic compact (and one-sided) jet ($v>0.6c$) during the
low/hard state \cite{Stirling01}, transient relativistic jets 
($v\geq 0.3c$) during state transitions \cite{Fender06b}, whereas no radio emission
is detected during the high/soft state. 

Arc-minute extended radio emission around 
Cygnus X-1 was found \cite{Marti96} using the VLA. Their disposition reminded 
of an elliptical ring-like shell with Cygnus X-1 offset from the center. 
Later, as reported in \cite{Gallo05}, 
such structure was recognised as a
jet-blown ring around Cygnus~X-1. 
This ring could be the result of a strong shock
that develops at the location where the pressure exerted by the collimated jet,
detected at milliarcsec scales, is balanced by the ISM \cite{Gallo05}. The observed bremsstrahlung radiation would be produced by the ionized gas behind the bow shock.

The instrument COMPTEL, on board the {\it CGRO}, did detect Cygnus~X-1 in the
1--30~MeV range several times \cite{Mcconnell00}. Unfortunately, EGRET
performed rather few observations of Cygnus~X-1 and did no detect the source.
Only a quite loose upper limit to the $\sim$100~MeV flux from Cygnus~X-1 is
available \cite{Hartman99}.

\begin{figure}
\resizebox{0.40\textwidth}{!}
   {\includegraphics{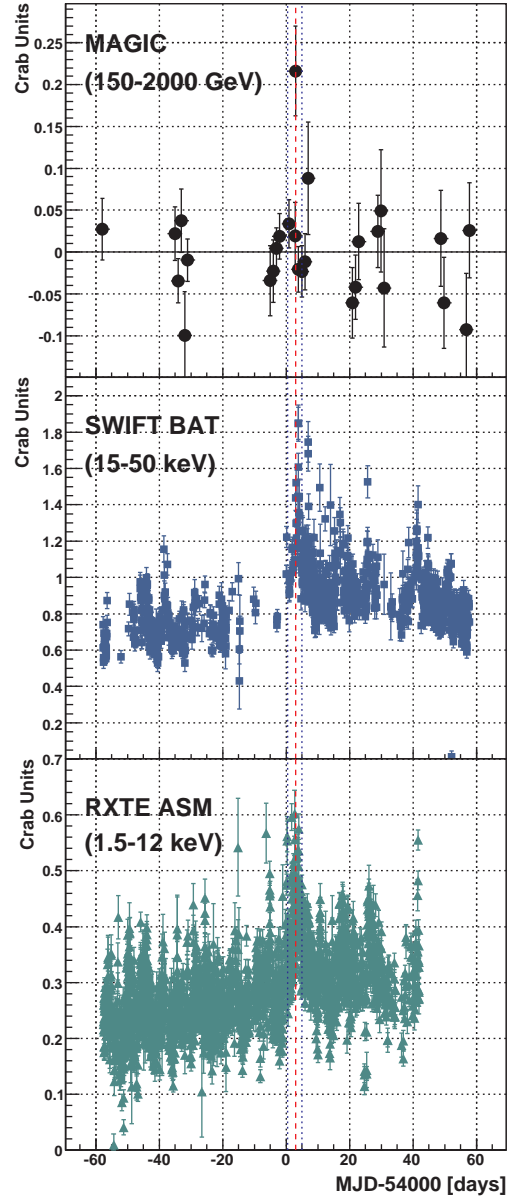}}
  \caption{MAGIC (top), {\it Swift}/BAT (middle) and {\it RXTE}/ASM (bottom) measured 
  fluxes from Cygnus~X-1 as a function of time \cite{Albert07}. 
The minor thicks in the X-axis are separated five days.}
\label{cygx1}
\end{figure}

\subsubsection{Fast TeV emission}

MAGIC observed Cygnus~X-1 in 2006 obtaining evidence of $\gamma$-ray signal 
with a significance of 4.9 $\sigma$ (4.1 $\sigma$ after trial correction)
\cite{Albert07}. The signal was variable and extending in a short time 
interval ($\sim$ 80 minutes), as can be seen in Fig.~\ref{cygx1}.
The measured excess is compatible with a point-like source at the position of 
Cygnus X-1 and excludes the nearby radio nebula powered by its relativistic jet.
 
The differential energy spectrum is well fitted by a power law given
as $dN/(dA dt dE) = (2.3\pm0.6)\times10^{-12}(E/$1 TeV)$^{-3.2\pm0.6}$.
The measured excess was observed at phase 0.91, very near the inferior 
conjunction of the optical companion and superior conjunction of the compact 
object \cite{Gies03}. According to existing models dealing with photon-photon 
absorption and cascading, there should not be detectable TeV emission at these 
orbital phases if its origin is close to the compact object \cite{Bednarek07}\cite{Bosch-Ramon08b} 
(behind the companion star in this configuration).

Interestingly, this detection occurred at the time when a flare was 
detected \cite{Turler06} at
hard X-rays by {\it INTEGRAL} (at a level of about 1.5~Crab (20--40~keV) and
1.8~Crab (40--80~keV)) and {\it Swift}/BAT, and at soft X-rays by {\it RXTE}
(see Fig.~\ref{cygx1}). The MAGIC detection occurred during a
particularly bright and hard X-ray flare that took place within a prolonged
low/hard state of Cygnus~X-1. The TeV peak appears to
precede a hard X-ray peak, while there is no particular change in soft X-rays.

\section{Flare TeV emission?}

Among the four TeV binaries detected up to now, there are two, LS~5039 and 
LS I +61 303, which present periodic TeV emission. The period of the VHE 
emission is coincident with the orbital period in the case of LS~5039 and very 
near to the orbital period in the case of LS I +61 303. Likely PSR~B1259$-$63 is 
also periodic at TeV, but its long (years) orbital period has not allowed   
the gathering of enough observational data to show it. The steady emission of Cygnus~X-1
has not yet been detected. 

In addition to the steady and periodic emission it seems that a new kind of
phenomenology could be present in some of these systems. The detected TeV emission 
in Cygnus~X-1 comes from a flaring episode and possibly this kind of fast emission
is common in other systems. 
In the
case of \lsi, in addition to the periodic TeV emission with a maximum at 
phase 0.6, it presents a flaring activity peaking at phase $\sim$0.8. This 
flaring emission is evident in the December 2006 run carried out by MAGIC (see 
panel labeled MJD=54070.7 in Fig.~\ref{lsi-MAGIC}).
{\it Swift}/XRT observed also \lsi\ along several orbital cycles and among them the 
December 2006 cycle, labeled orbit 5 in Figure 3 of \cite{Esposito07}. In 
this case there is also a temporal coincidence between the TeV and the X-ray 
flare. The other source that presents flaring TeV emission superposed
to the periodic-regular light-curve is LS~5039. As can be seen in 
Fig.~\ref{ls}, there is apparent flare activity around phase 0.8.

\section{Scenarios}

Two main models for the radio and high energy emission were developped after the association of LS~I +61~303 with
CG/2CG~135+01.  One suggests that the radio outbursts of LS~I~+61~303 are produced by streams of relativistic particles 
powered by episodes of accretion onto a compact object in a highly eccentric orbit, embedded in the mass outflow from the
companion B-star (\cite{Taylor82}, \cite{Taylor92}). Alternatively, LS~I~+61~303 might contain a non-accreting young pulsar
in orbit around a mass-losing B-star powered by the pulsar wind \cite{Maraschi81}. The pulsar model received a strong suport
after the discovery of PSR B1259$-$63 \cite{Tavani94}. However, the discovery of resolved radio structures, interpreted as
jets, pointed towards the microquasar scenario \cite{Massi01}. Several models based in this scenario were developped, being
leptonic (\cite{Bosch-Ramon04}, \cite{Bosch-Ramon06}, \cite{Gupta06}, \cite{Bednarek06}) or hadronic (\cite{Romero05},
\cite{Orellana07}).  These jet-like features have been reported several times, showing a puzzling behavior
\cite{Massi04}. Recently, VLBI observations show a rotating jet-like structure \cite{Dhawan06}. The changes of the 
radio morphology have been interpreted as produced by the interaction of the relativistic wind from a young pulsar
with the wind from its stellar companion, the scenario accepted for PSR B1259$-$63. In certain pulsar models (e.g.
\cite{Dubus06}), a cometary-like nebula of radio emitting particles would be expected. This structure would rotate along the
orbit pointing away from the companion star. The interaction of the relativistic wind from a young pulsar with the wind from
its stellar companion has been considered a viable scenario for explaining the observations of LS~5039 (in addition to PSR
B1259$-$63 and LS~I +61~303) (\cite{Dubus06}, \cite{Dubus08}). It has been noted that hydrodynamical simulations of pulsar/star wind
interactions (\cite{Romero07}, \cite{Bogovalov08}) do not show the elongated shape seen in the VLBI radio images of LS~I
+61~303, previously cited as strong evidence in favor of a pulsar/star wind interaction scenario.

A feature at very high energies that would allow the distinction between the accretion and the pulsar scenarios is a
line-type energy spectrum formed by the Comptonization of stellar photons by a mono-energetic pulsar wind, as shown for
PSR~B1259$-$63/SS2883 by \cite{Khangulyan07}. This has been also calculated for the cases of LS~I~+61~303 and LS~5039 by
\cite{Cerutti08}, although electromagnetic cascades were not accounted for, which was otherwise done by \cite{Sierpowska07}
in the case of LS~5039. 

Another interesting fact is that Cygnus~X-1 and LS~5039 show TeV emission around the superior conjunction of the compact
object, when the largest gamma-ray opacities are expected. To investigate the implications of these detections, given the
role of the magnetic field for the occurrance of electromagnetic cascading in these systems, the absorbed luminosity due to
pair creation in the stellar photon field for different emitter positions has been computed \cite{Bosch-Ramon08}. The
results suggest that the TeV emitters in Cygnus~X-1 and LS~5039 are located at a distance $> 10^{12}$ cm from the compact
object.
This would disfavor those models for which the emitter is well inside the system, like
the innermost-jet region (microquasar scenario), or the region between the pulsar and the primary star (standard pulsar
scenario) \cite{Bosch-Ramon08}. Similar results concerning the location of the emitter in the case of LS~5039 were already
discussed by \cite{Khangulyan08} based on
acceleration efficiency arguments. 



\begin{theacknowledgments}
The author acknowledges support of the Spanish Ministerio de Educaci\'on y 
Ciencia (MEC) under grant AYA2007-68034-C03-01 and FEDER funds. Fig~\cite{McConnell02}, taken from 2002 ApJ 572, 984 (DOI:10.1086/340436),  has been reproduced by permission of the AAS. I am indebted to Valent\'{\i} Bosch-Ramon,  Josep Mart\'{\i} and Marc Rib\'o for a careful
reading of the manuscript and their valuable comments. This research has made use
of the NASA's Astrophysics Data System Abstract Service, and of the SIMBAD database, operated at
CDS, Strasbourg, France.
\end{theacknowledgments}



\bibliographystyle{aipproc}   


\begin{thebibliography}{99}

\bibitem{Corbel02}
S.~Corbel, R.~P.~Fender, A.~K.~Tzioumis, {\it et al.}, \emph{Science} \textbf{298}, 196--199 (2002).

\bibitem{Aharonian05}
F.~A.~Aharonian, {\it et al.}, \emph{Science} \textbf{309}, 746-- 749 (2005).

\bibitem{Albert06}
J~Albert, {\it et al.}, \emph {Science} \textbf{312}, 1771--1773 (2006).

\bibitem{Aharonian05a}
F.~A.~Aharonian, {\it et al.}, \emph{Astron. Astrophys.} \textbf{442},
1-- 10 {2005}.

\bibitem{Liu06} 
Q.~Z.~Liu, J.~van~Paradijs, and E.~P.~J.~van den Heuvel, \emph{Astron. Astrophys.} \textbf{455}, 1165--1168 (2006).

\bibitem{Liu07} 
Q.~Z.~Liu, J.~van~Paradijs, and E.~P.~J.~van den Heuvel, \emph{Astron. Astrophys.} \textbf{469}, 807-- 810(2007).

\bibitem{Fender99}
R.~P.~Fender, M.~M.~Hanson, and G.~G~Pooley,  \emph{MNRAS} \textbf{308}, 
473--484 (1999).

\bibitem{Gregory72}
P.~C.~Gregory, P.~P.~Kronberg, E.~R.~Seaquist, {\it et al.}, \emph{Nat. 
Phys. Sci.} \textbf{239}, 114 (1972).

\bibitem{Marti01} 
J.~Mart\'{\i}, J.~M~Paredes, J. M., and M.~Peracaula, \emph{Astron. Astrophys.} \textbf{375}, 476--484 (2001).

\bibitem{Miller-Jones04}
J.~C.~A.~Miller-Jones, K.~M.~Blundell, M.~P.~Rupen, {\it et al.},  \emph{Astrophys. J.} \textbf{600}, 368--389 (2004).

\bibitem{Sanchez08}
J.~R.~S\'anchez-Sutil, J.~Mart{\'{\i}}, J.~A.~Combi, {\it et al.}, \emph{Astron. Astrophys.} \textbf{479}, 523--528 (2008).

\bibitem{Szostek08}
A.~Szostek, A~A.~Zdziarski, and M~L~McCollough, 
\emph{MNRAS} \textbf{388}, 1001--1010 (2008).

\bibitem{Trushkin08}
S.~A.~Trushkin, N.~N.~Nizhelskij, and J.~V.~Sotnikova, \emph{\em ATel} 
\textbf{1483}, (2008).

\bibitem{McConnell02}
M.~L.~McConnell, {\it et al.},  \emph{Astrophys. J.} \textbf{572}, 984--995 (2002). 

\bibitem{Zdziarski04} 
A.~A.~Zdziarski, and M.~Gierli{\'n}ski, \emph{PThPS} \textbf{155}, 99--119 (2004).

\bibitem{Giuliani08}
A.~ Giuliani,  {\it et al.}, 5th Science {\it AGILE} Workshop, 12--13 June 2008, Tor-Vergata, ESRIN.

\bibitem{Chadwick85}
P.~Chadwick, {\it et al.}, \emph{Nature} \textbf{318}, 642--644 (1985).

\bibitem{Hartman99} 
R.~C.~Hartman, {\it et al.}, \emph{Astrophys. J. Sup. Ser.} \textbf{123}, 79--202 (1999).

\bibitem{Tavani97}
M.~Tavani, and J.~Arons, \emph{Astrophys. J.} \textbf{477}, 439--464 (1997).

\bibitem{Neronov07}
A.~Neronov, and M.~Chernyakova, \emph{Ap\&SS} \textbf{309}, 253--259 (2007). 

\bibitem{Khangulyan07}
D.~Khangulyan, S.~Hnatic, F.~Aharonian, and S.~Bogovalov, \emph{MNRAS} 
\textbf{380}, 320--330 (2007). 

\bibitem{Kerschhaggl08}
M.~Kerschhaggl, {\it et al.}, in 4th Heidelberg International Symposium on High Energy Gamma-Ray Astronom, July 7-11, 2008, Heidelberg, Germany.

\bibitem{Taylor82}
A.~R.~Taylor, and P.~C.~Gregory, \emph{Astrophys. J.} \textbf{255}, 
210--216 (1982).

\bibitem{Gregory02}
P.~C.~Gregory, \emph{Astrophys. J.} \textbf{575}, 427--434 (2002).

\bibitem{Casares05a} 
J.~Casares, {\it et al.}, \emph {MNRAS} \textbf{360}, 1105--1109 (2005).

\bibitem{Frail91}
D.~ A.~ Frail, and R.~ M.~Hjellming, \emph{AJ} \textbf{101}, 2126--2130 (1991).

\bibitem{Massi04}
M.~ Massi, {\it et al.}, \emph{Astron. Astrophys.} \textbf{414}, L1--L4 (2004).

\bibitem{Dhawan06} 
V.~Dhawan, A.~Mioduszewski, and M.~Rupen, 2006,
in Proc. of the VI Microquasar Workshop, Como-2006.

\bibitem{Dubus06}  
G.~Dubus, \emph{Astron. Astrophys.} \textbf{456}, 801--817 (2006).

\bibitem{Paredes07}
J.~M.~Paredes, M., Rib{\'o}, V., Bosch-Ramon, J.~R., West, Y.~M., Butt, D.~F., 
Torres, and J.~Mart{\'{\i}}, \emph{Astrophys. J.} \textbf{664}, L39--L42 (2007).

\bibitem{Hermsen77}
W.~Hermsen, {\it et al.},  \emph{Nature} \textbf{269}, 494--495 (1977).

\bibitem{Gregory78}
P.~C.~Gregory, and A.~ R.~Taylor, \emph{Nature} \textbf{272}, 704--706 (1978).

\bibitem{Kniffen97}
D.~ A.~ Kniffen, {\it et al.}, \emph{Astrophys. J.} \textbf{486}, 126--131 (1997).

\bibitem{Schonfelder00}
V.~Sch\"onfelder, {\it et al.}, \emph {Astron. Astrophys. Suppl.} \textbf{143}, 145--179 (2000).

\bibitem{Sidro07}
N.~ Sidro, V.~Bosch-Ramon, Cortina, J., {\it et al.} (for the MAGIC Collaboration)
2007, Proc. of the 30th International Cosmic Ray Conference (M\'erida), in
press.

\bibitem{Acciari08}
V.~ A.~Acciari, {\it et al.}, \emph {Astrophys. J.} \textbf{679}, 1427--1432 (2008).

\bibitem{Paredes87}
J.~M.~Paredes 1987, PhD Thesis, Universitat de Barcelona.

\bibitem{Hutchings81}
J.~B.~Hutchings, and D.~Crampton, \emph{PASP} \textbf{93}, 486--489 (1981).

\bibitem{Mendelson89}
H.~Mendelson, and T.~Mazeh,  \emph{MNRAS} \textbf{239}, 733--740 (1989).

\bibitem{Mendelson94}
H.~Mendelson, and T.~Mazeh,  \emph{MNRAS} \textbf{267}, 1--4 (1994).

\bibitem{Paredes94}
J.~M.~Paredes, {\it et al.}, \emph{Astron. Astrophys.} \textbf{288}, 519--528 (1994).

\bibitem{Paredes97} 
J.~M.~Paredes, {\it et al.}, \emph{Astron. Astrophys.} \textbf{320}, 
L25--L28 (1997).

\bibitem{Zamanov99}
R.~ K.~Zamanov, J.~Mart\'{\i}, J.~M.~Paredes, {\it et~al.}, \emph{Astron. Astrophys.} \textbf{351}, 543--550 (1999).

\bibitem{Albert08}
J~Albert, {\it et al.}, \emph {Astrophys. J.}, (astro-ph:0806.1865)(2008)

\bibitem{Sidro08}
N.~ Sidro 2008, PhD Thesis, Universitat Aut\`onoma de Barcelona

\bibitem{Tavani98}
M.~Tavani, {\it et al.},  \emph{Astrophys. J.} \textbf{497}, L89--L91 (1998).

\bibitem{Motch97}
C.~Motch, {\it et al.}, \emph{Astron. Astrophys.} \textbf{323}, 853--875 (1997).

\bibitem{Marti98}
J.~Mart\'{\i}, J.~M.~Paredes, and M.~Rib\'o, \emph{Astron. Astrophys.} \textbf{338}, L71--L74 (1998).

\bibitem{Ribo99}
M.~Rib\'o, {\it et al.}, \emph{Astron. Astrophys.} \textbf{347}, 518--523 (1999).

\bibitem{Ribo02}
M.~Rib\'o, {\it et al.}, \emph{Astron. Astrophys.} \textbf{384}, 954--964 (2002).

\bibitem{Paredes00}
J.~M.~Paredes, J. Mart\'{\i}, M. Rib\'o, and M. Massi,  
\emph{Science} \textbf{288}, 2340--2342 (2000).

\bibitem{Clark01}
J.~S.~Clark, {\it et al.}, \emph{Astron. Astrophys.} \textbf{376}, 476--483 (2001).

\bibitem{McSwain01}
M.~V.~McSwain, {\it et al.}, \emph{Astrophys. J.} \textbf{558}, L43--L46 (2001).

\bibitem{Casares05b} 
J.~Casares, {\it et al.}, \emph {MNRAS} \textbf{364}, 899--908 (2005).

\bibitem{Collmar04}
W.~Collmar, in Proc. of the 4th {\it AGILE} Science Workshop, pp.177--182, Frascati
(Rome) on 11-13 June 2003. Eds M.~Tavani, A.~Pellizzoni, and S. Vercellone. (2004).

\bibitem{Aharonian06}
F.~A.~Aharonian, {\it et al.}, \emph{Astron. Astrophys.} \textbf{460}, 743--749 (2006).

\bibitem{Boettcher 2007}
M.~B\"ottcher, \emph{Astroparticle Physics} \textbf{27}, 278--285 (2007).

\bibitem{McSwain02}
M.~V.~McSwain, and D.~R.~Gies, \emph{Astrophys. J.} \textbf{568}, L27--L30 (2002).

\bibitem{Moldon07}
J. Moldón, Master thesis, Universitat de Barcelona, 2007. 

\bibitem{Gies86} 
D.~R.~Gies, and C.~T.~Bolton, \emph{Astrophys. J.} \textbf{304}, 371--393 (1986).

\bibitem{Ziolkowski05} 
J.~Zi\'o{\l}kowski, \emph{MNRAS} \textbf{358}, 851--859 (2005).

\bibitem{Wilms06} 
J.~Wilms, \emph{Astron. Astrophys.} \textbf{447}, 245--261 (2006).

\bibitem{Fender04} 
R.~Fender, {\it et al.}, \emph{MNRAS} \textbf{355}, 1105----1118 (2004).

\bibitem{Stirling01} 
A.~M.~Stirling, {\it et al.}, \emph{MNRAS} \textbf{327}, 1273--1278 (2001).

\bibitem{Fender06b} 
R.~Fender, {\it et al.}, \emph{MNRAS} \textbf{369}, 603--607 (2006).

\bibitem{Marti96} 
J.~Mart\'{\i}, {\it et al.}, \emph{Astron. Astrophys.} \textbf{306}, 449--454 (1996).

\bibitem{Gallo05} 
E.~Gallo, {\it et al.}, \emph{Nature} \textbf{436}, 819--821 (2005).

\bibitem{Mcconnell00} 
M.~L.~McConnell, {\it et al.}, \emph{Astrophys. J.} \textbf{543}, 928--937 (2000).

\bibitem{Albert07} 
J.~Albert, {\it et al.}, \emph{Astrophys. J.} \textbf{665}, L51--L54 (2007).

\bibitem{Gies03} 
D.~R.~Gies, {\it et al.}, \emph{Astrophys. J.} \textbf{583}, 424--436 (2003).

\bibitem{Bednarek07} 
W.~Bednarek, and F.~Giovannelli, \emph{Astron. Astrophys.} \textbf{464}, 
437--445 (2007).

\bibitem{Bosch-Ramon08b}
V.~Bosch-Ramon, in Proc. First La Plata International School on Astronomy and Geophysics, 2008, in press (astro-ph/0805.1707).

\bibitem{Turler06} 
M.~T\"urler, {\it et al.}, \emph{\em ATel} \textbf{911}, (2006).

\bibitem{Ribo08}
M.~Rib\'o, {\it et al.}, \emph{Astron. Astrophys.} \textbf{481}, 17--20 (2008).

\bibitem{Ribo02t}
M.~Rib\'o 2002, PhD Thesis, Universitat de Barcelona. 

\bibitem{Esposito07}
P.~Esposito, {\it et al.}, \emph{Astron. Astrophys.} \textbf{474}, 575--578 (2007).

\bibitem{Taylor92}
A.~R.~Taylor, {\it et al.},  \emph{Astrophys. J.} \textbf{395}, 268--274 (1992).

\bibitem{Maraschi81}
L.~Maraschi, and A.~ Treves, \emph{MNRAS} \textbf{194}, 1--5 (1981).

\bibitem{Tavani94}
M.~Tavani, J.~Arons, and V.~M.~Kaspi,  \emph{Astrophys. J.} \textbf{433}, L37--L40 (1994). 

\bibitem{Massi01}
M.~Massi, M.~Rib\'o, J.~M.~Paredes, M.~Peracaula, and R.~Estalella, R., \emph{Astron. Astrophys.} \textbf{376}, 217--223 (2001).

\bibitem{Bosch-Ramon04}
V.~Bosch-Ramon, and J.~M.~Paredes, \emph{Astron. Astrophys.} \textbf{425}, 1069--1074 (2004).

\bibitem{Bosch-Ramon06}
V.~Bosch-Ramon, J.~M.~Paredes, G.~R.~Romero, and M.~Rib\'o, \emph{Astron. Astrophys.} \textbf{459}, L25--L28 (2006).

\bibitem{Gupta06}
S.~Gupta, and M.~B\"ottcher, \emph{Astrophys. J.} \textbf{650}, L123--L126 (2006).

\bibitem{Bednarek06}
W.~Bednarek, \emph {MNRAS} \textbf{371}, 1737--1743 (2006).

\bibitem{Romero05}
G.~E.~Romero, H.~R.~Christianen, and M.~Orellana, \emph{Astrophys. J.} \textbf{632}, 1093--1098 (2005).

\bibitem{Orellana07}
M.~Orellana, and G.~E.~Romero, \emph{ApSS} \textbf{309}, 333--338 (2007).

\bibitem{Dubus08}  
G.~Dubus, B.~Cerutti, and G.~Henri, \emph{Astron. Astrophys.} \textbf{477}, 691--700 (2008).

\bibitem{Romero07}
G.~E.~Romero, {\it et al.}, \emph{Astron. Astrophys.} \textbf{474}, 15--22 (2007).

\bibitem{Bogovalov08}
S.~V.~Bogovalov, D.~Khangulyan, A.~V.~Koldoba, {\it et al.}, \emph {MNRAS} \textbf{387}, 63--72 (2008). 

\bibitem{Cerutti08}
B.~Cerutti, G.~Dubus, and G.~Henri, \emph {Astron. Astrophys.} in press 
(arxiv: 0807.1226)(2008).

\bibitem{Sierpowska07}
A.~Sierpowska-Bartosik, and D.~F.~Torres, \emph{Astrophys. J.} \textbf{671}, L145--L148 (2007).

\bibitem{Bosch-Ramon08}
V.~Bosch-Ramon, D.~Khangulyan1, and F.~A.~Aharonian, \emph{Astron. Astrophys.} {\bf in press} (astro-ph:0808.1540)

\bibitem{Khangulyan08}
D.~Khangulyan, F.~Aharonian, and V.~Bosch-Ramon, \emph{MNRAS} 
\textbf{383}, 467--478 (2008). 







\end{thebibliography}

\IfFileExists{\jobname.bbl}{}
 {\typeout{}
  \typeout{******************************************}
  \typeout{** Please run "bibtex \jobname" to optain}
  \typeout{** the bibliography and then re-run LaTeX}
  \typeout{** twice to fix the references!}
  \typeout{******************************************}
  \typeout{}
 }

\end{document}